\documentclass[epj,final]{svjour}
\usepackage{graphicx}
\usepackage{epsfig}
\usepackage{amsmath,amssymb}
\usepackage{color}

\begin{document}

\authorrunning{P.~M. Dinh et al}
\title{Dynamics of cluster deposition on Ar surface}
\author{ P.~M. Dinh\inst{1}  \and F. Fehrer\inst{2}
\and P.-G. Reinhard\inst{2}  \and E. Suraud\inst{1}
} 
\institute{ 
Laboratoire de Physique Th\'eorique, UMR 5152, Universit\'e P. Sabatier,
F-31062 Toulouse cedex, France 
\and 
Institut f{\"u}r Theoretische Physik, Universit{\"a}t Erlangen,
Staudtstrasse 7, D-91058 Erlangen, Germany
}
\date{\today / Received: date / Revised version: date}
\abstract{ 
Using a combined quantum mechanical/classical method, we study the
dynamics of deposition of small Na clusters on Ar(001) surface.  We
work out basic mechanisms by systematic variation of substrate
activity, impact energy, cluster orientations, cluster sizes, and
charges. The soft Ar material is found to serve as an extremely
efficient shock absorber which provides cluster capture in a broad
range of impact energies. Reflection is only observed in combination
with destruction of the substrate.  The kinetic energy of the
impinging cluster is rapidly transfered at first impact.  The
distribution of the collision energy over the substrate proceeds very
fast with velocity of sound. The full thermalization of ionic and
atomic energies goes at a much slower pace with times of several ps.
Charged clusters are found to have a much stronger interface
interaction and thus get in significantly closer contact with the
surface.
} 
\PACS{ 
 { 36.40.Gk, 36.40.Mr,36.40.Sx, 36.40.Vz, 61.46.Bc}
 {Atomic and molecular clusters} 
} 

\maketitle

\section{Introduction}

Clusters on surfaces are an appealing area of research which have
motivated numerous studies \cite{Bru00}.  The literature on the topic
is flourishing and many conferences have been (fully or partly)
devoted to these
researches, see for example the series of recent ISSPIC conferences
\cite{ISSPIC9,ISSPIC10,ISSPIC11,ISSPIC12}.  It is now possible to make
a direct deposition of size selected clusters on a substrate
\cite{Bin01,Har00}. This opens up new possibilities for the synthesis
of nano-structured surfaces. But the deposition process is not
necessarily simple and it may lead to a significant modification of
the cluster, both in terms of its electronic structure and of its
ionic geometry.  There is a subtle interplay of the interface
energy, electronic band structure of the substrate, and surface
corrugation.  These various aspects have already been investigated in
great detail, especially from the structural point of view, both
experimentally \cite{Exp1,Exp2,Exp3} and theoretically
\cite{BL,CL,HBL,MH,Koh96,Koh97a,Koh00,Ipa02}.
The situation becomes even more involved when one considers the
deposition dynamics itself. But that makes the case also more
interesting. And in a such involved situation, one should start to
disentangle the various influences by studying simple systems, i.e.
simple geometries and material combinations. In this contribution, we
are presenting first results for the deposition dynamics of small Na
clusters on Ar(001) surface. To that end, we employ a hierarchical
model which treats the different subsystems at different levels of
refinement, depending on their relevance for the whole process.

While the presence of a substrate makes the experimental handling of
clusters easier, it strongly complicates the theoretical description
because of the huge number of degrees of freedom of the surface.  Most
theoretical approaches thus 
remain limited to molecular dynamics (MD).  The MD provides
the cheapest way to access, at least in a gross way, the dynamics of
the substrate. However, its applicability remains restricted to a
narrow range of physical situations, because a proper description of
electronic degrees of freedom is missing.  It is nevertheless 
crucial to try to
account for the surface electronic degrees of freedom, in particular
if non-adiabatic processes become involved.

A first simplification is to use relatively simple cluster/substrate
combinations as, e.g., simple metal cluster on an insulator surface.
Because of its inert nature the insulator surface can be included at a
lower level of description. This was, e.g., explored for the case of
Na clusters on NaCl in \cite{Ipa02,Ipa04} where the cluster electrons
were described fully quantum-mechanically while the substrate is
frozen and just serves to deliver an effective interface potential
(itself tuned to {\it ab-initio} calculations \cite{MH}).
However  rare-gas atoms show sizeable polarization response.
These effects need to be included.  A somewhat better description of
surface degrees of freedom can be achieved by allowing for a minimum
of dynamical response of the substrate, as recently proposed for the
description of Na clusters embedded in rare gases
\cite{Ger04b,Feh06a,Feh05c}. The electronic degrees of freedom of the
cluster were treated microscopically by time-dependent
density-functional theory while Na ions and substrate atoms were
handled with classical MD, taking carefully into account the dynamical
polarization of the atoms.
This model belongs to the family of coupled Quantum-Mechanical with
Molecular-Mechanical methods (QM/MM) which are often used in other
fields as, e.g., bio-chemistry \cite{Fie90a,Gao96a,Gre96a} or surface
physics \cite{Mit93a,Nas01a}.
%


In this paper we apply this hierarchical approach to study 
the deposition of small Na clusters on an Ar surface. 
The paper is organized as follows: 
Section  \ref{sec:model} provides a brief presentation of the
hierarchical approach and the construction of a surface.
In section \ref{sec:surf}, we discuss the effects of  the
dynamical treatment of  the Ar atoms.
In section \ref{sec:energ}, we discuss variation of collisional energy
in terms of ionic motion  and energetic observables.
In section \ref{sec:geom}, we study the influence of the
initial collisional geometry.
And in section \ref{sec:charge}, we look at 
variation of cluster size and charge.

\section{Model}
\label{sec:model}

We give here a short summary of the model, for more details see
appendix \ref{sec:enfundetail} and \cite{Feh05a}.
The Na cluster is described in the TDLDA-MD approach, which was well
validated for linear and non-linear dynamics of free metal clusters
\cite{Rei03a,Cal00}. The Na electrons are treated by means of
density-functional theory at the level of the time-dependent
local-density approximation (TDLDA), and the Na$^+$ ions are
propagated by means of molecular dynamics (MD). The electron-ion
interaction is described by soft, local pseudo-potentials
\cite{Kue99}.  Each Ar atom has two classical degrees of freedom:
center-of-mass and electrical dipole moment. The dipoles allow an
explicit treatment of the dynamical polarizability of the atoms.  The
key constituent of the dynamic cluster-atom interaction are
polarization potentials \cite{Dic58}.  A local pseudo-potential is
added for the electron-Ar short-range repulsion, modeled following
\cite{Dup96}, and with a final slight readjustment to the NaAr dimer
as benchmark (bond length, binding energy, and optical excitation).
The Na-Ar Van-der-Waals interaction is computed via the variance of
dipole operators \cite{Ger04b,Feh05a,Dup96}.  The atom-atom
interactions are described by a standard Lennard-Jones potential,
while the Ar-Na$^+$ subsystem is treated by means of effective
potentials from the literature \cite{Rez95}. More details on these 
various components, including the choice of parameters can be found 
in appendix.

The total energy thus composed is the starting point for variation leading
to the time-dependent Kohn-Sham equations for the cluster electrons
coupled with Hamiltonian equations of motion for the classical degrees
of freedom (Na$^+$ ions, Ar positions and dipoles). The initial
condition is obtained from solving the corresponding stationary
equations and finally boosting the Na cluster to the
wanted impact energy.

The numerical solution proceeds with standard methods \cite{Cal00}.
We use space-grid techniques to solve the (time-dependent) Kohn-Sham
equations for the cluster electrons.  The time propagation is based on
a time-splitting method, and the stationary solution is attained by
accelerated gradient iterations.  The electronic mean-field is treated
in axially averaged approximation \cite{Mon94a,Mon95a} which was found
to provide an acceptable approximation for the present test case of
the nearly axial Na$_6$ cluster impinging on a surface.  In the
following, the symmetry axis will be denoted by the $z$ axis.  It also 
corresponds to the deposition axis. The
Na$^+$ ions and the Ar atoms are, of course, treated in full three
dimensions.

Several observables can be computed. We employ here simple geometric
and energetic indicators, the ionic/atomic positions and their kinetic
energies.  We have checked also electronic observables, as dipole
oscillations and ionization. Both play a minor role for the 
analysis (e.g., electronic emission during deposition is less than
0.02~\%).  However, this does not imply that the 
dynamical path evolves along a simple and geometrically
prescribed Born-Oppenheimer surface.
Cluster electrons and ions couple to the many atoms in the 
surface through short range collision and long-range
dynamical polarizability which excites all degrees of freedom
in a state far from equilibrium.

Our main test case is Na$_6$ on Ar surface.  Na$_6$ consists out of
five Na ions in a ring plus one 6th ion topping the ring.  The top ion
sits on the symmetry axis.  The Ar(001) surface is
modeled as six layers of 8$\times$8 squares containing together 384
Ar atoms. The squares are copied periodically in both horizontal
directions to simulate an infinite surface. To stabilize the
underlying (supposedly infinite) crystal structure, the atomic
positions in the lower two layers are frozen at the bulk positions.
%
{Some comparisons have been performed with the larger 
{substrate configuration}
Ar$_{512}$, composed by 6 active layers and 2 frozen ones. We have
found good agreement, justifying that 4 active layers are
sufficient in the present dynamical regime.}
Finally in our energetic analysis, the Ar atoms in the vicinity of the
impact point will be emphasized. This includes an hemisphere around the
impact point (4$\times$4 in first layer, 3$\times$3 in second, and
2$\times$2 in third).  We will see that this impact hemisphere
carries almost the whole energy transfer in the deposition process
(cf. right panel of Fig.~\ref{fig:na6ekint}).

\section{The effect of the surface}
\label{sec:surf}

In a first run, we investigate the influence of the modeling of the
Ar substrate.  The Na$_6$ center-of-mass starts from 15 $a_0$ above
the surface with an initial velocity in the $z$ direction,
corresponding to a kinetic energy $E^0_{\rm kin}$ of 0.06 Ry.
Experimentally, this setup should lead to sticking, if one
extrapolates available experimental data with proper scaling laws (see
the discussion below) \cite{Har90}.
\begin{figure}[htbp]
\centerline{\epsfig{figure=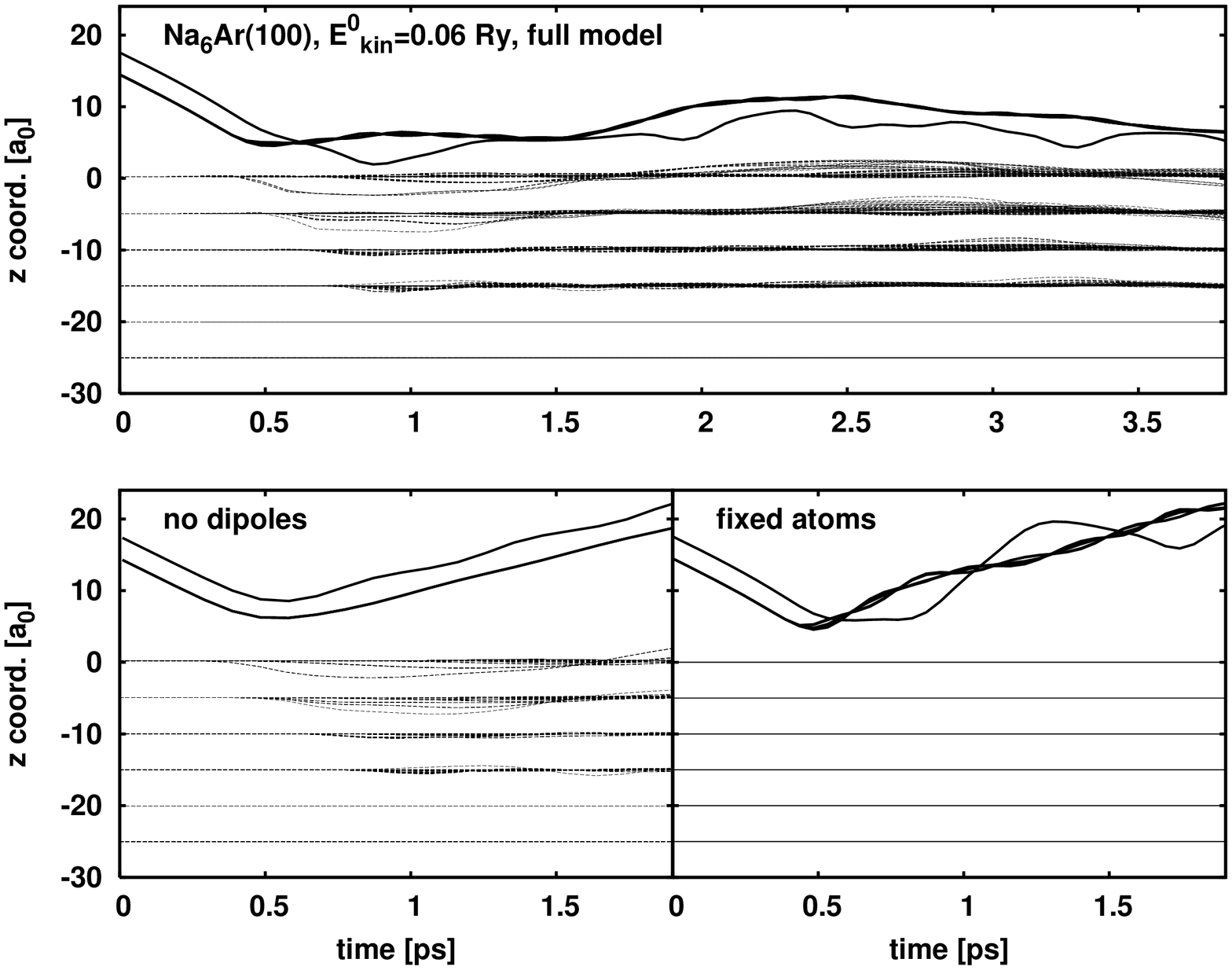,width=8.7cm}}
\caption{\label{fig:bounce}
$z$ coordinates of the Na$_6$ cluster (full curves) deposited on Ar$_{384}$
(dotted) for an impact energy of 0.06 Ry.
Three different levels of treatment are shown.
Upper panel: full model.
Lower left panel: dynamical dipoles of Ar atoms switched off.
Lower right panel: totally frozen Ar atoms.
} 
\end{figure}
In order to analyze the impact of the various ingredients of the
model, Fig.~\ref{fig:bounce} compares the deposition dynamics for
three different levels of treatment, full model, dynamical dipoles
switched off, and {frozen atomic positions (while maintaining
dynamical dipoles).}
The cluster is initially
in ``bottom'' configuration where the pentagon is closer to the
surface and the top ion faces away.
In all three cases, the first 500 fs of the dynamics proceed similar
showing the cluster steadily approaching the surface. Then the
collision and the subsequent evolution proceed very differently. 
For the fixed Ar atoms (lower right panel), the surface reacts rigidly
and one observes an immediate reflection of Na$_6$ combined with
strong internal excitations as oscillations of the top ion through the
pentagon.
For mobile Ar atoms without dipole dynamics (lower left panel),
there is again reflection, but now with a visible amount of excitation
energy going to the substrate while less is left for internal motion
of the cluster. 
For the case of fully active Ar atoms (upper panel),
the substrate absorbs most of the cluster's kinetic energy and the
cluster remains tied to the surface, with 
its  distance to the surface performing 
large oscillations in the attractive potential well
created by the Ar dipoles.  We thus observe a relatively soft
landing of the Na$_6$ in "bottom" configuration.  Soft means here
that the surface and the cluster are still moderately perturbed.
The overall comparison  of the three calculations 
demonstrates the importance of a full dynamical 
treatment of the Ar surface, going beyond a 
mere Molecular Dynamics of Ar positions but also 
accounting for their polarizabilities through time-dependent dipoles. 

The perturbation produced by the impact of the cluster on the first
layer propagates straightforwardly through the substrate as a sound
wave with about the speed of sound inside Ar bulk, namely around 20-30
a$_0$/ps. When this sound wave reaches the 5th (and frozen) layer, it
is reflected upwards. When coming back to the surface it transfers 
some residual momentum to the cluster. However that momentum transfer
remains 
moderate and leaves the cluster in a captured state.  That back-flow of
momentum would not show up in that pronounced manner for an infinitely
deep material. We have checked that in this energy range and in the case 
of the deposit of a single Na atom, adding 2 more Ar layers. This indeed 
does not change the scenario qualitatively as the 2 added active Ar shells 
do absorb  a rather limited amount of kinetic 
energy. It might nevertheless modify {\em in fine} the deposition
threshold quantitatively. However the model with frozen bottom layers
has a realistic touch. It provides a zeroth-order simulation of surfaces
from hard materials covered by a thin layer of rare gas
\cite{Ira05}. Mind nevertheless that the case may differ  
at a quantitative level. In the case of Ar coated metals, for example, 
the underlying metal provides an extra attraction on the deposited
cluster. This may thus modify the details of the deposition  scenario,
especially at the side of energetic considerations and deposition
thresholds. 

Altogether, Fig.~\ref{fig:bounce} demonstrates the crucial role
played by the elasticity of the surface.  Ar is extremely soft and
serves as a true stopper material for gentle deposition. The proper
dynamical treatment of the surface also appears as essential.

\section{Energetic analysis}
\label{sec:energ}

\begin{figure*}
\begin{center}
\epsfig{figure=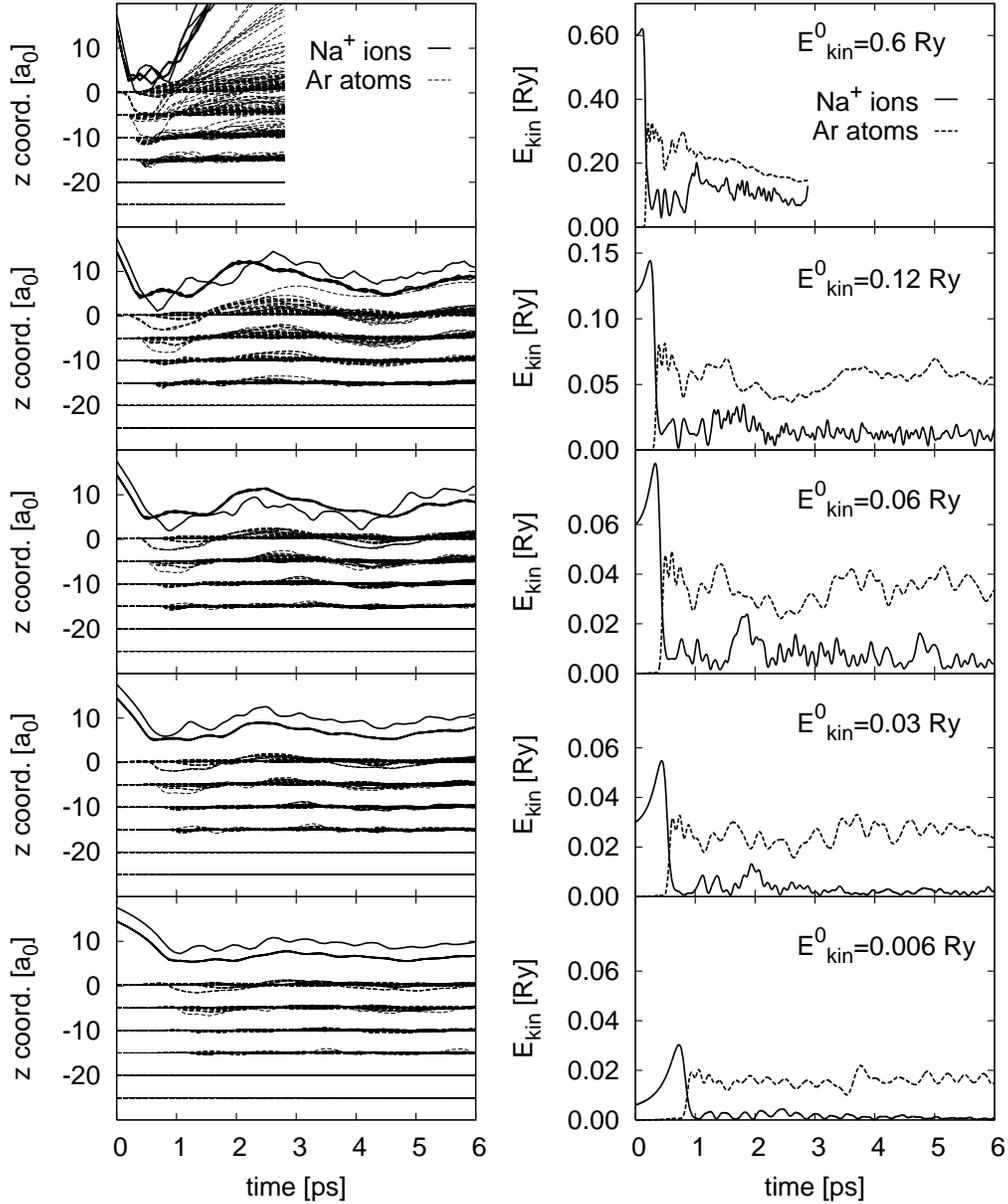,width=14cm}
\caption{\label{fig:na6ekine} 
Time evolution of deposition of Na$_6$ on Ar$_{384}$ for five
different initial kinetic energies $E^0_{\rm kin}$ as indicated.  
Left column: $z$ coordinates of Na$^+$ ions (heavy solid line)
and of Ar atoms (faint dashed).
Right column: kinetic
energies for the Na cluster (solid lines) and for the Ar
system (dashed).
}
\end{center}
\end{figure*}
\begin{figure*}
\begin{center}
\epsfig{figure=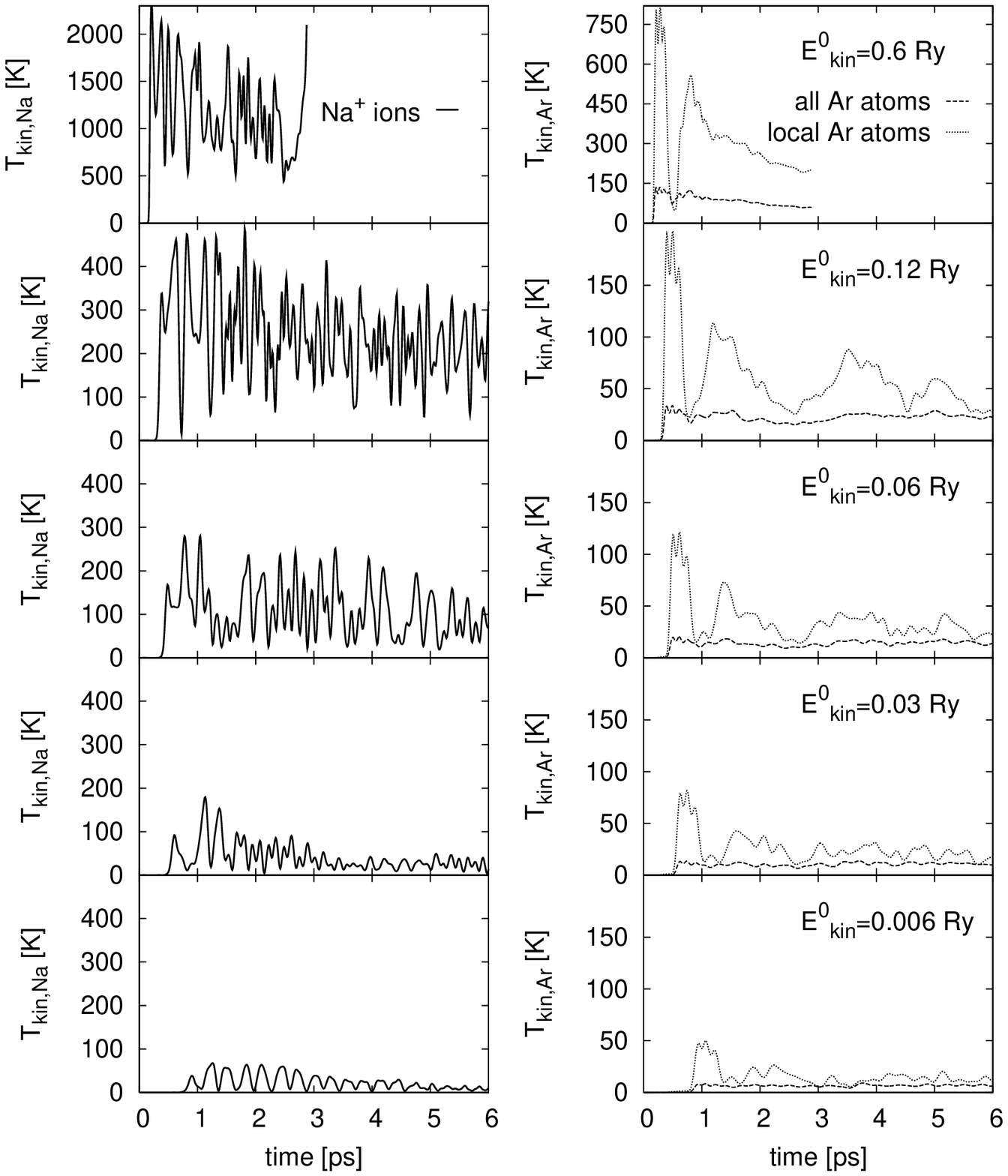,width=13cm}
\caption{\label{fig:na6ekint} 
Time evolution of deposition of Na$_6$ on Ar$_{384}$ for five
different initial kinetic energies $E^0_{\rm kin}$ as indicated
(Figure \ref{fig:na6ekine} continued).  
Left column:  kinetic temperature of Na$^+$ ions.
Right column: kinetic temperature of the Ar substrate
taken over the impact hemisphere (dashed)
denoted as ``local'' and  over all remaining atoms (dotted) 
denoted ``all''.
}
\end{center}
\end{figure*}
In this section, we study the dependence on the initial kinetic energy
$E^0_{\rm kin}$ of Na$_6$. This amounts to change its
initial velocity along the $z$ direction. 
Fig.~\ref{fig:na6ekine} summarizes results for the time
evolution viewed through spatial coordinates and kinetic energies.
  A very broad range of
initial kinetic energies $E^0_{\rm kin}$, from 0.6 Ry down to 0.006
Ry, is considered.
The detailed atomic and ionic $z$ coordinates (left column) show at
first glance an overall similarity of all the different cases, except
for the highest energy. There is a very fast stopping and quick
capture of the Na cluster followed by persistent oscillations of the
cluster distance plus some internal oscillations. The cluster momentum
transferred at impact propagates as a sound wave through the
substrate. The perturbation of the substrate, of course, increases
with the initial energy. The case of the highest energy differs. One
sees reflection of the cluster, however at the price of severe
destruction of the surface. It corroborates the view that Ar is an
extremely efficient stopper material.  
{At this point, a comparison with experiments performed on
 comparable systems~\cite{Har90} can be made.}
The measurements of~\cite{Har90} were{, in fact,} 
carried out with the much
heavier Ag material but once deposition energies are properly scaled,
{our results are in qualitative agreement.}
Note that sufficiently high deposition energies were not attained in
the experiments of~\cite{Har90} to access  the destructive regime. 
{In our simulations, the threshold for destruction of the
substrate seems to lie between 0.05 and 0.1 Ry per metal atom.}
It should finally be noted that for all cases of
non-destructive deposition, the course of the process does not depend
much on the initial projectile velocity. The same result has been
found in studies of deposition on large, but finite, Ar
clusters~\cite{Din07a}.

The kinetic energies are shown in the second column of
Fig.~\ref{fig:na6ekine}.
The kinetic energy of the Na cluster increases before contact due to
the long-range polarization interaction which is attractive. The
additional acceleration depends on the initial velocity. Slower
velocities allow for more energy gain, in relative value, 
 since the cluster moves for a
longer time in the attractive regime. The kinetic energy is multiplied
almost by a factor 8 for the lowest $E^0_{\rm kin}=0.006$ Ry whereas
only insignificant acceleration occurs for the fastest collision.
At time of impact, there emerges a very fast and almost complete
energy transfer from the Na cluster to the substrate. In less than 0.5
ps, the Ar carry away almost the whole Na cluster kinetic energy and
only very little residual kinetic energy is left to the Na
cluster. Longer times are needed for the final relaxation
processes. Note also the revival of the cluster kinetic energy a few
ps after impact. This is due to the come back of the reflected 
wave in the substrate, as discussed above.

It is instructive to analyze energy transfers in terms of
temperature. To that end, we calculate the intrinsic kinetic energy
$E^{\rm int}_{\rm kin}$ by subtracting the contribution from the
center-of-mass motion which is particularly relevant for the Na
cluster. The kinetic temperature is then defined as 
$T_{\rm kin} = 2 E^{\rm int}_{\rm kin}/3N$,
where $N=6$ for the cluster and 384 for
the substrate. The time evolution of the kinetic temperature is 
plotted in Fig.~\ref{fig:na6ekint}.  The first column shows $T_{\rm
kin}$ for the cluster.  The initial phase is purely center-of-mass
motion without intrinsic excitation. The temperature jumps at the
moment of impact due to the large perturbation of all constituents in
the impact zone. The jump ends close to the final temperature and
there remains a slow and moderate relaxation to thermal equilibrium.
The right column of Fig.~\ref{fig:na6ekint} shows the
temperatures for the substrate in two ways, taken over the atoms in
the vicinity of the impact point (the ``impact hemisphere'' as
explained at the end of section \ref{sec:model}) or over all Ar atoms
outside the impact hemisphere.  The differences between these both
temperatures are huge. The initial energy transfer goes preferably to
the region around the impact point.  The kinetic temperature of the
impact hemisphere shows recurrent bumps. They are related to slow
oscillations within the substrate (see first column of 
Fig.~\ref{fig:na6ekine}) and associated
energy exchange between potential and kinetic energy. The average
temperature relaxes slowly to that of the total system.  The
amplitude of the oscillations and the relaxation time strongly depend
on the initial energy, ranging from about 5 ps for the weakly excited
cases to outside our simulation time for the heftier processes.  Note
that the average temperature for the highest energy is above the
melting point of about 84 K~\cite{Wah12,Pol64}, 
indicating once more the destruction of
the substrate.
The electronic excitation during the collision amounts to
small dipole oscillation with an amplitude of about 0.05 a$_0$ and
related energy content of about 6.8 meV.

\begin{figure}[htbp]
\centerline{\epsfig{figure=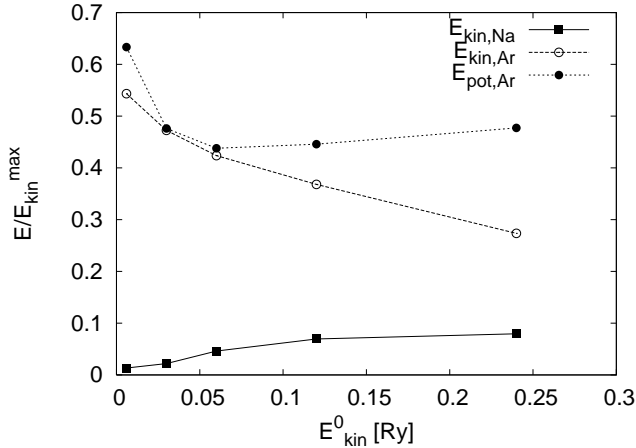,width=8.7cm}}
\caption{\label{fig:ekinratio}
Partitioning of final energies over cluster kinetic energy,
substrate kinetic, and substrate potential energy
normalized to initial energy.
}
\end{figure}

In the second column of Fig.~\ref{fig:na6ekine}, we had tracked the
kinetic energy transfer from the cluster to the surface as a function
of time. It is also interesting to study the final repartition of the
initial energy. For that purpose, the values of the ionic kinetic, the
atomic potential and the atomic kinetic energies after 6 ps are
recorded and are normalized to the maximum kinetic energy, ${E_{\rm
kin}}^{\rm max}$, reached before impact. The thus obtained  energy ratios are
plotted as function of $E^0_{\rm kin}$ in Fig.~\ref{fig:ekinratio}.
The figure shows that 
the energy share left for the Na cluster increases with increasing
deposition energy. Nonetheless, the energy loss at the side of the Na cluster is
dramatic, even for the most violent case.  For the Ar substrate, we
see an equal share between potential and kinetic energies, except for
the highest initial energy. The gain in potential energy is related to the
moderate spatial rearrangements in the Ar substrate. For $E^0_{\rm
kin}=0.6$ Ry, the kinetic energy is 2/3 times larger than the
potential one. Indeed the emitted Ar in this case are blown away with
a substantial velocity (see upper left panel of
Fig.~\ref{fig:na6ekine}).

\section{Geometrical effects}
\label{sec:geom}

\begin{figure*}
\centerline{\epsfig{figure=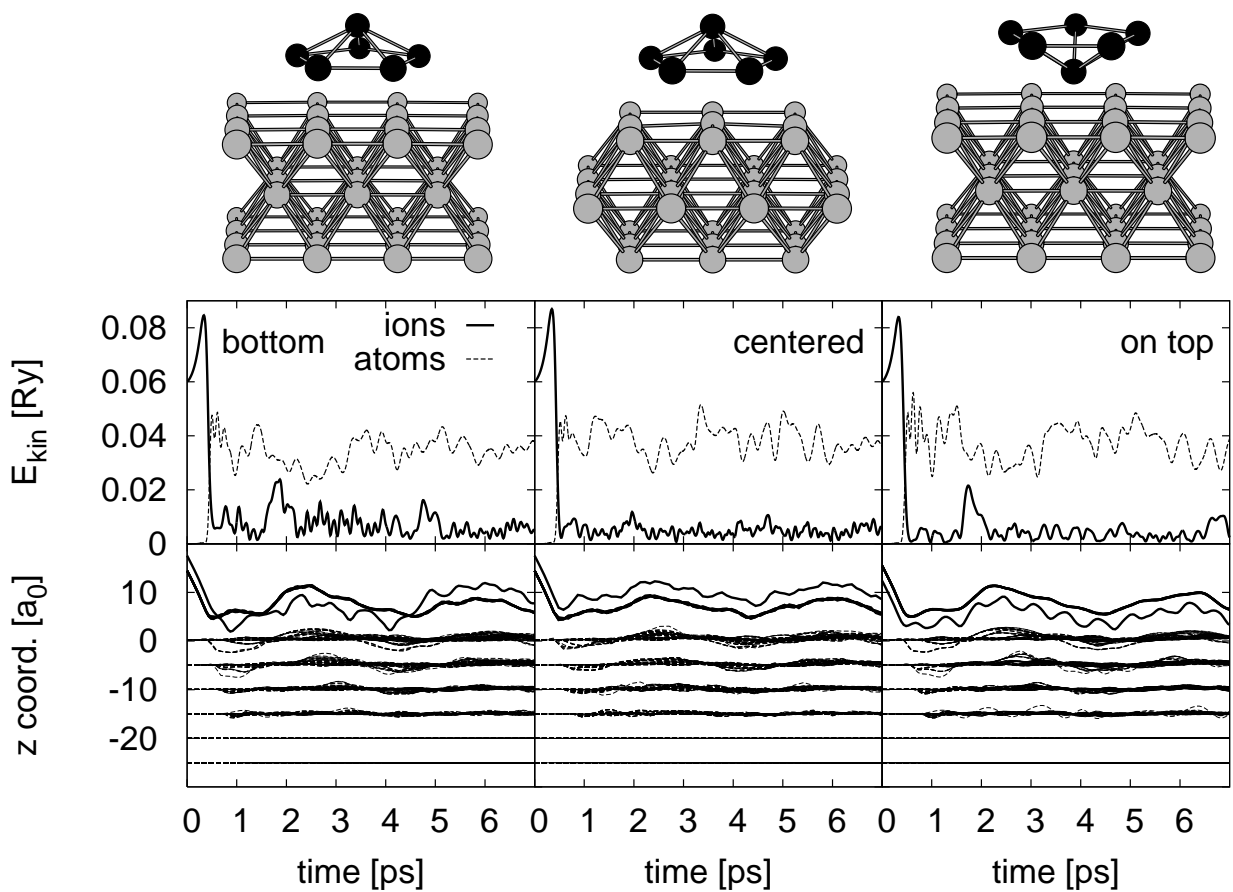,width=11.7cm,angle=90}}
\caption{\label{fig:na6geom}
Time evolution of $z$ coordinates (lower panels) and kinetic energies
(middle panels) for three different collision geometries as sketched
in the upper most panels. All cases start with initial kinetic energy
of 0.06 Ry and correspond to a deposition of Na$_6$ on Ar$_{384}$.
} 
\end{figure*}

There are different possible initial geometries (orientation, position
relative to surface) which may produce different reaction pattern (for
an example, see the case of Na$_6$ on NaCl \cite{Ipa02,Ipa04}).
All results
reported above are obtained from the starting configuration displayed
on the left column of Fig.~\ref{fig:na6geom}, denoted by ``bottom'':
the top ion faces away from the surface and the cluster center axis
is placed above an interstitial position of the first Ar layer. Two
other configurations have been studied. The ``centered'' one (middle
column of Fig.~\ref{fig:na6geom}) is similar to the ``bottom''
configuration but with the axis exactly above an Ar atom of the first
layer. The ``on top'' configuration (right column of
Fig.~\ref{fig:na6geom}) is obtained from the ``bottom'' configuration
by reversing the top ion to face towards the surface such that the top
ion hits the Ar surface first. In all cases, the cluster
center-of-mass is initially positioned 15 $a_0$ above the first layer.

The results for the time evolution of the $z$ coordinates and the kinetic
energies are shown in Fig.~\ref{fig:na6geom}.  At first glance, all
three results look very similar.  
Interesting differences become apparent in the details. The average
kinetic energy of the cluster is largest for the ``bottom''
configuration and smallest for ``on top''. This relates to the amount
of Ar core repulsion presented to the cluster. For ``bottom'', the
axis lies on an interstitial position and the Na ions stay closer to
the sites of the Ar atoms. For ``centered'', the most repulsive site
points in between the Na ions of the pentagon, the interaction
 being that way weakened.
The situation  is somehow similar 
for ``on top'' where the now closest top ion dives
into the interstitial site with maximum distance to the atoms.
This also explains why the top ion does not go
through the pentagon in that case. 
There is also a slight difference in the final distance to the
surface.  The ``on top'' configuration seems to stay closer which may
be, again, due to a minimization of core repulsion in this configuration.

\section{Size and charge effects}
\label{sec:charge}

\begin{figure*}
\centerline{
\epsfig{figure=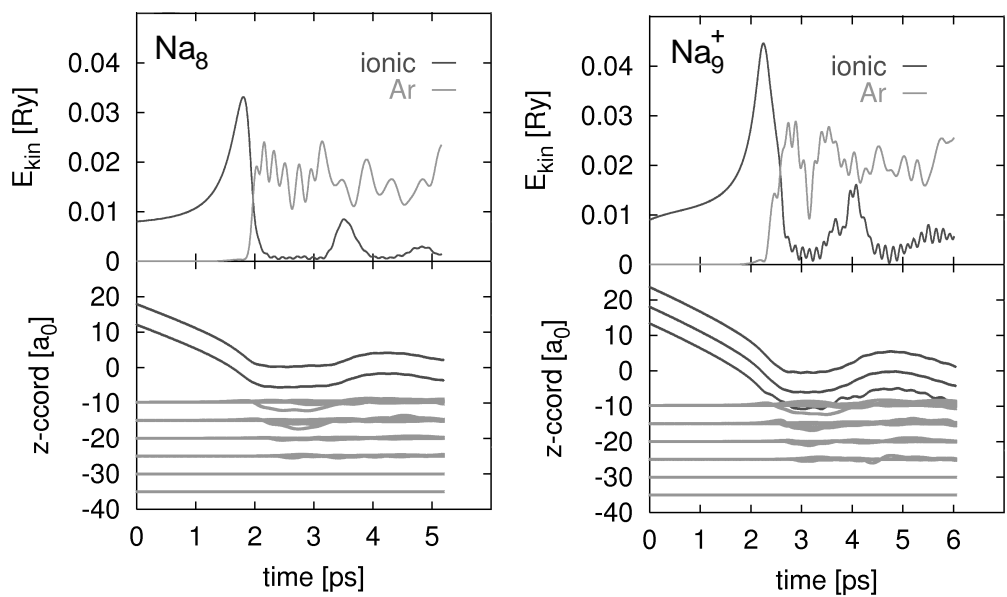,width=14cm}
}
\caption{\label{fig:na8ar384}
Time evolution of deposition for Na$_8$ (left part) and
Na$_9^+$ on Ar$_{384}$ surface at impact energy of
0.001 Ry per Na ion. Shown are the $z$ coordinates (lower) as well as
the kinetic energies of Na ions and Ar atoms (upper). 
}
\end{figure*}

In the previous sections, we finally observed in most cases the soft
landing of the Na$_6$ on the Ar surface. One may suspect that this is
favored by the oblate shape of this cluster which provides a large
contact area. It is interesting to test the deposition scenario for
different cluster geometries.  A good alternative is Na$_8$. As is
well known the 8 valence electrons of Na$_8$ form a closed shell
\cite{Hee93,Bra93} (magic number of electrons), which leads to a close
to spherical (up to the impact of ions) electronic shape. The overall
shape of Na$_8$ is, in turn, itself close to spherical and is thus
significantly different from the especially oblate one of Na$_6$. %
The left part of Fig.~\ref{fig:na8ar384} shows the time evolution for
Na$_8$ impinging on Ar$_{384}$ surface.  The scenario for Na$_8$ is
very similar to what we had observed for Na$_6$, with efficient
stopping, sudden energy transfer, and final capture. One even observes
less excitation and deformation of the Na$_8$ cluster than in the
Na$_6$ cases. This effect is probably due to the fact that Na$_8$ is
electronically magic and thus especially stable.

The right part of Fig.~\ref{fig:na8ar384} shows the result for the
charged 
cluster Na$_9^+$. It has also the magic electron number 8 and is near to
spherical shape.  While the overall bouncing scenario is qualitatively
similar (with sound wave bounce), the net result is quantitatively
different.  Differences show up already in the initial phase. The
finite net charge yields stronger attractive polarization interaction
with the substrate and thus acquires much more acceleration. The thus
higher impact energy and the stronger interaction produce a more
violent collision. Most remarkable is the closer attachment to the
surface which, again, is due to the much stronger polarization
interaction. The bottom of the Na$_9^+$ cluster is even merging into
the surface layer. This small example indicates that the deposition
of charged clusters is a most interesting topic which deserves further
studies, also in direct relation with experiments \cite{Har90,Sie06}.

\section{Conclusions}

In this paper, we have investigated the deposition dynamics of 
small Na clusters on an Ar(001) surface. To that end, we have employed a
hierarchical model treating the Na cluster in full detail by
time-dependent density-functional theory, the Na$^+$ ions by classical
dynamics, and the Ar atoms also at a classical level with position
and dipole polarization as dynamical variables. We have studied in
detail the influence of initial conditions on the deposition process,
varying substrate activity, impact energy, collisional geometry, and
size as well as charge of the Na cluster.

We have found that the gross features of deposition dynamics are much
the same in all cases. The Ar system acts always as a very efficient
and smooth stopper which absorbs almost all (more than 90\%) of the
impact energy and thus yields capture over a broad range of energies.
Cluster reflection, which is typical for hard surfaces, can  be forced
only at such high impact energies that the surface is destroyed.
The impinging Na cluster is stopped at first impact and its energy is
transferred to the Ar system in less than 0.5 ps.  The transfer is
almost complete leaving less than 10\% of the energy 
to the cluster. The transferred
energy is distributed also very quickly over all Ar atoms, propagating
like a sound wave with speed of sound through the medium.  The full
thermalization over all ionic and atomic degrees of freedom proceeds
on a slow scale with time of 5 ps and more (increasing with increasing
impact energy).
The collisional geometry (initial orientation and horizontal position
relative to the surface) has no effect on the overall dynamical
pattern, but determines subtle details of the captured state as,
e.g., remaining cluster energy or distance to the surface.
Cluster size and shape seem to make little difference. Cluster charge
however can 
change a lot. It enhances overall excitation as well as binding of the
cluster to the surface and produces a much tighter final configuration.

\bigskip

Acknowledgments: This work was supported by the DFG, project nr. RE
322/10-1, the French-German exchange program PROCOPE nr. 07523TE, the
CNRS Programme ``Mat\'eriaux'' (CPR-ISMIR), Institut Universitaire de
France, the Humboldt foundation and a Gay-Lussac price.

\appendix

\section{The Na-Ar energy functional in detail}
\label{sec:enfundetail}

The degrees of freedom of the model are the wavefunctions of valence
electrons of the metal cluster, $\{\varphi_n({\bf r}),n=1...N_{\rm el}\}$,
the coordinates of the cluster's  Na$^+$ ion cores, 
$\{{\bf R}_I,I=1...N_{\rm ion}\}$, of the Ar atoms cores Ar$^{Q+}$),
$\{{\bf R}_a,a=1...N_{\rm Ar}\}$, and of the Ar valence clouds, 
$\{{\bf R'}_a,a=1...N_{\rm Ar}\}$.
>From the given total energy, the corresponding equations of motion
are derived in a standard manner by variation. This leads to the
(time-dependent) Kohn-Sham equations for the one single-particle
wavefunctions $\varphi_n({\bf r})$ of the cluster electrons, and
Hamiltonian equations of motion for the other three degrees of
freedom, thus treated by classical mole\-cular dynamics (MD). 
For the valence cluster electrons, we use
a density functional theory at the level of the time-dependent
local-density approximation (TDLDA), augmented with an average-density
self-interaction correction (ADSIC)~\cite{Leg02}. The density of these
electrons is given naturally as defined in mean-field theories and
reads $\rho_{\rm el}({\bf r}) = \sum_n\left|\varphi_n^{\mbox{}}({\bf
  r})\right|^2$.
An Ar atom is
described by two constituents with opposite charge, positive Ar core
and negative Ar valence cloud, which allows a correct description of
polarization dynamics. In order to avoid singularities, we associate a
smooth (Gaussian) charge distribution to 
both constituents having width $\sigma_{\rm Ar}$ of the order of the p
shell "size" in Ar atoms, in the spirit of~\cite{Dup96}:
\begin{equation} 
\begin{split}
  \rho_{{\rm Ar},a}({\bf r}) &= 
  \frac{e \, Q}{\pi^{3/2}_{\mbox{}}\sigma_{\rm Ar}^3} \times \cr
  & \hspace{-1cm} \times 
  \Big[
   \exp{\left(-\frac{({\bf r}\!-\!{\bf R}^{\mbox{}}_a)^2}{\sigma_{\rm
  Ar}^2}\right)} 
   -
   \exp{\left(-\frac{({\bf r}\!-\!{\bf R}'_a)^2}
                    {\sigma_{\rm Ar}^2}\right)}
  \Big].
\label{eq:Ardistri}
\end{split}
\end{equation}
The corresponding Coulomb potential exerted by the Ar atoms is
related to the charge distribution (\ref{eq:Ardistri}) by the Poisson
equation, and reads:
\begin{equation} 
\begin{split}
  V^{\rm(pol)}_{{\rm Ar},a}({\bf r})
  =
  e^2{Q^{\mbox{}}} 
  \Big[
  &\frac{\mbox{erf}\left(|{\bf r}\!-\!{\bf R}^{\mbox{}}_a|
          /\sigma_{\rm Ar}^{\mbox{}}\right)}
        {|{\bf r}\!-\!{\bf R}^{\mbox{}}_a|} \cr
  & -
   \frac{\mbox{erf}\left(|{\bf r} \!-\! {\bf R}'_a|/\sigma_{\rm
  Ar}^{\mbox{}}\right)}  {|{\bf r}\!-\!{\bf R}'_a|}
  \Big],
\label{eq:Arpolpot}
\end{split}
\end{equation}
where \mbox{$\mbox{erf}(r) = \frac{2}{\sqrt{\pi}}\int_0^r \textrm
  dx\,e^{-x^2}$} stands for the error function.
As for the Na$^+$ ions, their dynamical polarizability is
neglected and we treat them simply as charged point particles.

The total energy of the system is composed as:
\begin{equation}
  E_{\rm total}
  =
  E_{\rm Na cluster}
  +
  E_{\rm Ar}
  +
  E_{\rm coupl}
  +
  E_{\rm VdW}
  \quad.
\label{eq:Etot}
\end{equation}
The energy of the Na cluster $E_{\rm Na cluster}$ consists out of
TDLDA (with SIC) for the electrons, MD for ions, and a coupling of
both by soft, local pseudo-potentials, for details
see~\cite{Kue99,Cal00,Rei03a}. 
The Ar system and its coupling to the cluster are described by
\begin{eqnarray}
\label{eq:E_Ar}
  E_{\rm Ar}
  &=&
  \sum_a \frac{{\bf P}_a^2}{2M_{\rm Ar}} 
  +
  \sum_a \frac{{{\bf P}'_{a}}^2}{2m_{\rm Ar}}
  +
  \frac{1}{2} k_{\rm Ar}\left({\bf R}'_{a}\!-\!{\bf R}_{a}\right)^2 \cr
  && \hspace{-1.2cm} +
  \sum_{a<a'}
  \left[
    \int \textrm d{\bf r} \, \rho_{{\rm Ar},a}({\bf r})
    V^{\rm(pol)}_{{\rm Ar},a'}({\bf r})
    +
    V^{\rm(core)}_{\rm Ar,Ar}({\bf R}_a \!-\! {\bf R}_{a'})
  \right],
\\
\label{eq:E_coupl}
  E_{\rm coupl}
  &=&
  \sum_{I,a}\left[
    V^{\rm(pol)}_{{\rm Ar},a}({\bf R}_{I})
    +
    V'_{\rm Na,Ar}({\bf R}_I \!-\! {\bf R}_a)
  \right] \cr
  && \hspace{-1.2cm}+
  \int \textrm d{\bf r}\rho_{\rm el}({\bf r})\sum_a \left[
    V^{\rm(pol)}_{{\rm Ar},a}({\bf r})
    +
    W_{\rm el,Ar}(|{\bf r} \!-\! {\bf R}_a|)
  \right],
\\
\label{eq:EvdW}
  E_{\rm VdW}
  &=&  
  \frac{e^2}{2} \sum_a \alpha_a
  \Big[
    \frac{
       \left(\int{\textrm d{\bf r} \, {\bf f}_a({\bf r}) \rho_{\rm
  el}({\bf r})}\right)^2 }{N_{\rm el}} \cr
  && \hspace{2cm}
    - \int{\textrm d{\bf r} \, {\bf f}_a({\bf r})^2 \rho_{\rm el}({\bf
  r})} 
  \Big] \;,
\\
\label{eq:fa}
  {\bf f}_a({\bf r})
  &=&
  \nabla\frac{\mbox{erf}\left(|{\bf r}\!-\!{\bf R}^{\mbox{}}_a|
          /\sigma_{\rm Ar}^{\mbox{}}\right)}
        {|{\bf r}\!-\!{\bf R}^{\mbox{}}_a|}
  \quad.
\end{eqnarray}

The Van der Waals interaction between cluster electrons and Ar
dipoles is written in Eq.~(\ref{eq:EvdW}) as a correlation from the
dipole excitation in the Ar atom coupled with a dipole excitation in
the cluster.
We exploit that the plasmon frequency $\omega_{\rm Mie}$ is far below
the excitations in the Ar atom.
This simplifies the term to the variance of the dipole operator in
the cluster, using the regularized dipole operator ${\bf f}_a$,
defined in Eq.(\ref{eq:fa}), corresponding to the smoothened Ar charge
distributions~\cite{Feh05c,Gro98}. The full dipole variance is
simplified in terms of the local variance. 

The interaction of one Ar atom with the other constituents (Ar
atoms, Na$^+$ ions, cluster electrons) results from the balance 
between a strong repulsive core potential that falls off exponentially
and an equally strong attraction from dipole polarizability. 
The (most important) polarization potential, $V^{\rm(pol)}_{{\rm
    Ar},a}$, is described by 
a valence electron cloud oscillating against the Ar core
ion. Its parameters are the effective charge of valence cloud $Q$, 
the effective mass of valence cloud $m_{\rm Ar}=Qm_{\rm el}$, 
the restoring force for dipoles $k_{\rm Ar}$, and 
the width of the core and valence clouds $\sigma_{\rm Ar}$.
The $Q$ and $k_{\rm Ar}$ are adjusted to reproduce experimental
data on dynamical
polarizability $\alpha_D(\omega)$ of the Ar atom at low frequencies,
namely the static limit 
$\alpha_D(\omega\!=\!0)$ and the second derivative
of $\alpha''_D(\omega\!''=\!0)$.
The width $\sigma_{\rm Ar}$ is determined consistently such that the
restoring force from the folded Coulomb force (for small
displacements) reproduces the spring constant $k_{\rm Ar}$.

The short range repulsion is provided by the various core potentials.
For the Ar-Ar core interaction in Eq.~(\ref{eq:E_Ar}), we employ a
Lennard-Jones type potential with parameters reproducing binding
properties of bulk Ar~: 
\begin{equation}
  V_{\rm Ar,Ar}^{\rm (core)}(R)
  = 
  e^2 A_{\rm Ar} \left[
  \left( R_{\rm Ar}/R \right)^{12}
 -\left( R_{\rm Ar}/R \right)^{6}
  \right].
\label{eq:VArAr}
\end{equation}
The Na-Ar core potential $V'_{\rm Na,Ar}$ in Eq.~(\ref{eq:E_coupl}) is
chosen according to~\cite{Rez95}, within 
properly avoiding double counting of the dipole
polarization-potential, hence the following form~:
\begin{equation}
\begin{split}
  V'_{\rm Na,Ar}(R)
  =
  e^2\Bigg[
  A_{\rm Na} & \frac{e^{-\beta_{\rm Na} R}}{R} \cr
  -\
  \frac{2}{1+e^{\alpha_{\rm Na}/R}}
  & \left( \frac{\alpha_{\rm Ar}}{2R^4} + \frac{C_{\rm Na,6}}{R^6} 
  + \frac{C_{\rm Na,8}}{R^8}\right)
  \Bigg] \cr
  + \, e^2\frac{\alpha_{\rm Ar}}{2R^3} 
  {\bf R} \cdot \nabla_{\bf R}&
  \frac{\mbox{erf}(R/\sqrt{2}\sigma_{\rm Ar})}{R} \quad.
\label{eq:VpArNa}
\end{split}
\end{equation}
Finally the pseudo-potential $W_{\rm el,Ar}$ in Eq.~(\ref{eq:E_coupl})
for the electron-Ar core repulsion has been modeled according to the
proposal of~\cite{Dup96}~:
\begin{equation}
  W_{\rm el,Ar}(r)
  =
  e^2\frac{A_{\rm el}}{1+e^{\beta_{\rm el}(r - r_{\rm el})}} \quad .
\label{eq:VArel}
\end{equation}

The various contributions are calibrated from independent
sources, with a final fine tuning to the NaAr dimer (bond length,
binding energy, and optical excitation spectrum)
modifying only the term $W_{\rm elAr}$. The parameters are
summarized in table~\ref{tab:params}. The third column of the table
indicates the source for the parameters. 

\begin{table*}
\begin{center}
\begin{tabular}{|l|l|l|}
\hline
 \rule[-8pt]{0pt}{22pt}
 $V^{\rm(pol)}_{{\rm Ar},a}$
&
 $q_{\rm Ar}
  =
  \frac{\alpha_{\rm Ar}m_{\rm el}\omega_0^2}{e^2}$
 \;,\;
 $k_{\rm Ar}
  =
  \frac{e^2q_{\rm Ar}^2}{\alpha_{\rm Ar}}$
 \;,\;
 $m_{\rm Ar}=q_{\rm Ar}m_{\rm el}$
&
 $\alpha_{\rm Ar}$=11.08$\,{\rm a}_0^3$
\\
 \rule[-12pt]{0pt}{22pt}
 &
 $\sigma_{\rm RG}
  =
  \left(\alpha_{\rm Ar}\frac{4\pi}{3(2\pi)^{3/2}}  \right)^{1/3}$
&
 \raisebox{12pt}{$\omega_0 = 1.755\,{\rm Ry}$}
\\
\hline
 \rule[-6pt]{0pt}{18pt}
 $W_{\rm elAr}$
&
 $A_{\rm el}$=0.47  
 \;,\;
 $\beta_{\rm el}$=1.6941\,/a$_0$  
 \;,\;
 $r_{\rm el}=$2.2 a$_0$ 
&
 fit to NaAr dimer~\cite{Gro98,Rho02a}
\\ 
\hline
 \rule[-8pt]{0pt}{22pt}
$V^{\rm(core)}_{\rm ArAr}$
&
 $A_{\rm Ar}$=$1.367*10^{-3}$ Ry
 \;,\;
 $R_{\rm Ar}$=6.501 a$_0$ 
&
fit to bulk Ar
\\
\hline
 \rule[-6pt]{0pt}{18pt}
 $V'_{\rm ArNa}$
&
 $\beta_{\rm Na}$= 1.7624 a$_0^{-1}$
 \;,\;
 $\alpha_{\rm Na}$= 1.815 a$_0$
 \;,\;
 $A_{\rm Na}$= 334.85 
&
\\
 \rule[-6pt]{0pt}{18pt}
&
 $C_{\rm{Na},6}$= 52.5 a$_0^6$
 \;,\;
 $C_{\rm Na,8}$= 1383 a$_0^8$
&
after \cite{Rez95}
\\
\hline
\end{tabular}
\end{center}
\caption{
Parameters for the various model potentials.
}
\label{tab:params}
\end{table*}

\bibliographystyle{epj}
\bibliography{cluster,add}
 
\end{document}